\documentclass[10pt,letterpaper]{article}

\usepackage{cogsci}
\usepackage{graphicx}
\cogscifinalcopy 

\usepackage[
  style=apa,
  natbib=true,
  annotation=false,
]{biblatex}
\usepackage{float} 

\title{Forgetting Our Way to Shared Meaning: Effects of Forgetting on Conceptual Alignment in a Non-Partnership Coordination Game}

\author[1]{\mbox{Landon C.L. Liu (lcliu@andrew.cmu.edu)}}
\author[3]{\mbox{Alan Tsang (	AlanTsang@cunet.carleton.ca)}}
\author[4]{\mbox{Mary Kelly (mary.kelly4@carleton.ca)}}
\affil[1]{Department of Philosophy, Carnegie Mellon University}
\affil[1]{Institute for Complex Social Dynamics, Carnegie Mellon University}
\affil[3]{Department of Computer Science, Carleton University}
\affil[4]{Department of Cognitive Science, Carleton University}

\begin{document}

\maketitle

\begin{abstract}
Shared meaning in language requires people to learn and agree on categories. We ask how characteristics of agents' memories change the emergence and evolution of shared meaning. Without a coordination game, models of conceptual semantics cannot explain how shared meaning emerges and changes in groups of people; however, existing games assume that players share payoffs in a partnership setting. We model conceptual alignment as a non-partnership game and illustrate differences in actual and perceived conceptual convergence from counterfactual simulations using agents with varying levels of adaptiveness and memory degradation. We found that adaptive players achieved actual convergence faster and had closer final conceptual regions than non-adaptive players, while non-adaptive players \textit{perceived} convergence earlier. Weighing novel information less over time resulted in more stable agreements than fixing the weight of novel information. Memory features are critical to the emergence and evolution of actual and perceived convergence.


\end{abstract}

\section{Introduction}
\enquote{How has your week been?} is often followed by, \enquote{It's been good.} Politeness aside, even with an earnest and forthcoming friend, what constitutes \enquote{a good week} can vary remarkably depending on the friend. The concept of \enquote{goodness} here, localized to just events that can possibly happen to a person in a given week, is not a concept that suffers from a lack of data. Hopefully, most of us are being asked this question on a regular basis. We have no shortage of relevant data from regular use. Yet, it can be difficult to confidently feel that we always understand what a good week means--for some, that means successfully evading devastating calamity. For others, it may require achieving brilliance. The concept of \enquote{a good week,} as opposed to, say, \enquote{a bad week,} can be thought of as a partition of the concept of a week, with some decision boundary between \textit{good} and \textit{bad.} What we are interested in most broadly is how partitions converge in similarity as shared meanings develop, and how shared meanings develop based cognitive features of our language learning capacities. We are interested in how features of semantic memory (i.e., interpretable as rates of forgetting, levels of entrenchment, or reduced/a continued reduction in plasticity) affect the quality, the stability, and the speed of our shared agreements on these partitions. What can an architecture that allows for semantic memory tell us about misalignment, premature agreement and fragmentation in real linguistic communities?

Semantic memory is a heavily studied topic, especially with the rise in the use of and the research into Large Language Models \citep[LLMs;][]{bubeck2023sparks,Vaswani2017transformer}; however, LLMs face critical theoretical and practical problems. First, LLMs face what has been called the symbol grounding problem \citep{harnad_symbol_1990}; it is not clear how meaning is represented or even extracted from the training datasets of LLMs, or that it is indeed \enquote{grounded}. Second, training LLMs is informationally expensive when compared to empirical observations of human learning. Humans, particularly young children, can learn new words and their meanings remarkably quickly; for example, children only need a few instances of a word to learn its meaning; far fewer than the large number of examples an LLM would need to \enquote{learn} the meaning of a word \citep{gardenfors_geometry_2014}. Although there is no consensus on the exact representations used by semantic memory in the human mind, there is a well-established tradition of modeling the semantics of words as high-dimensional vectors \citep[e.g.,][]{dumont_exploiting_2023,kumar_semantic_2021,Landauer1997}, which includes LLMs, but also includes other approaches, such as vector-symbolic architectures \citep{Jones2006,Kelly2020indirect,mannering2021catastrophic}.
\citet{lieto_conceptual_2017} suggests \citet{gardenfors_geometry_2014}'s Conceptual Semantics as a lingua franca to provide a theoretical underpinning for vector-symbolic architectures \citep{Gayler2003,Kleyko2022,Kleyko2023,Plate1995} and other vector-space representations of semantics. 

In \citet{gardenfors_geometry_2014}'s theory of conceptual semantics, meaning is represented as \enquote{conceptual regions} which are subregions in a Euclidean space representative of semantic memory in the mind. These regions are constrained to be convex regions formed through Voronoi tessellations from conceptual prototypes (i.e. idealized objects which represent concepts). Then, new objects can be immediately classified based on their similarity to these conceptual prototypes; new objects are assumed to belong to the concept which is formed by the closest prototypical object.\footnote{For example, one conceptual prototype of a dog may be a very common dog such as a Labrador or a Golden Retriever; another conceptual prototype of a cat may be an American Shorthair. If an agent has these prototypes, if they are shown a new furry animal (suppose these are the only furry animals they can conceive of existing), for example, what we understand to be wolf, they may be inclined to call it a dog.} From this representation, \citet{gardenfors_geometry_2014} proposes a semantic theory which he considers to have both grounded meaning (in response to \citet{harnad_symbol_1990}'s symbol grounding problem) as well as fast, human-like learning.

However, given that conceptual semantics is a (partial) theory of language, its theory should be able to also allow people to share meaning to allow for efficient communication. Toward this end, \citet{gardenfors_geometry_2014} proposes an alignment game as the solution to the program. He proposes that in real life, there are consequences from our daily lives that incentivize us to want to ensure that we are indeed referring to the same objects and concepts. For example, suppose that you have an allergy to animals which you refer to as \enquote{cats}, and that I have a pet in my home (but you are not sure if the pet is what you understand to be a cat.) If I invite you to my home for dinner, you might ask me if my pet is a cat. If I do not have the same approximate definition of what constitutes a cat, I may tell you that I have what I believe to be dog; if the pet is indeed what you would call a cat and would cause you to suffer an allergic reaction, I would suffer the embarrassment of not properly warning you take anti-histamines (and not knowing what a cat is.)

\noindent
One formation of this game has taken the form of a dyadic, sender-receiver game \citep[e.g.,][]{jager_evolution_2007,warglien_semantics_2013}. \citet{jager_evolution_2007} produces a sender receiver game, whereby the sender and receiver both have conceptual spaces, and separate functions mapping conceptual spaces to signals; they have a joint utility reflecting the similarity between the meaning in the speakers' mind and the  receivers' mind; their payoffs both depend on conceptual distances as measured in the space of the receivers' mind. This makes this game a partnership game, which have a set of Nash equilibria which are an evolutionarily stable set \citep{akin_recurrence_1982}. 

This paper's main aim is to study how the adaptivity of semantic memory affects the emergence and evolution of concepts. To do this, we propose a novel game to describe conceptual alignment; our game allows the 
\enquote{feeling of alignment} of each agent to depend on their own understanding of concepts. As our model is a non-symmetric, non-partnership game, our description of language alignment does not entail the strong evolutionary equilibrium results that characterize symmetric partnership games. We see two possible contributions from our framework. 

First, we hope that with weaker assumptions, we can begin studying when meaning does not converge across communities.\footnote{Although not presented, we have preliminary results that show trade-offs between convergence quality and network connectivity given time constraints.} Language has developed in such a way that even though we continue to be increasingly connected, there remain socioeconomic markers that form community differences in meaning: for example, there are tangible differences between the understanding of highly morally or politically loaded terms, such as \enquote{democratic,} \enquote{freedom,} \enquote{evil,} etc. Are there non-social, cognitive reasons for these differences? This paper explores the role that memory adaptivity plays on our perception of alignment and our actual alignment with each another. 

Second, our simple model can explore the consequences of varying assumptions regarding adaptability to shifts in meaning. This has consequences not just for understanding in humans, but also for the future development of LLMs. When we give models new information, how should they weigh this information relative to their priors? One possible solution is to continuously aggregate information and to weigh new information no more heavily than old information. As the model gains more information, new information becomes a proportionally smaller and smaller contribution to the model's overall information set. For example, if we were to retrain LLMs on datasets that are strictly expansions of prior datasets, we would be  producing lineages of LLMs that become increasingly non-adaptive. This is unlike what we see in humans and animals, which have a well-established preference for new information \citep{Bolhuis1988rats,greene1986sources}. However, if the training datasets grow exponentially between LLM updates, or are otherwise normalized to bias for recent data, retraining LLMs can be considered an adaptive update. This paper explores the tradeoffs between adaptive and non-adaptive strategies for language evolution and concept learning; depending on the desired use-cases for the LLM, there may be more appropriate memory update strategies.

\section{Theoretical Model \& Methodology}
\subsection{Base Model}
In our two-player game, both agents start randomly generated memories represented as sets of partitions of some $\mathbb{R}^n$ space. These subspaces are constructed by conceptual prototypes; a region of a conceptual prototype is the collection of points that are closest to the prototype centroid $c_i$. We assume that both players have the same ordering on prototypes (i.e. for every player, the region generated by $c_i$ is composed of values strictly less than the region of $c_j$ if $i < j$;) they just may vary on how close together or far apart they are. The decision boundary between some conceptual prototype $c_i$ and another $c_j$ is their halfway point. Players do not know the other's memory, or their update strategy. Every round, both players are shown a point in the subspace,\footnote{We assume perfect perception here for simplicity, although this can be easily relaxed.} then, they announce their classification (e.g. which prototype is the closest.)
Informally, players receive the highest payoff when their signals match their partner's. When their signals do not match, each player's payoff is equal to their perception of the distance between the object and the closest boundary of the concept that the other player has referenced. 

For example, suppose that Player 1 and 2 have three regions on the unit interval, $[A, B, C]$, as specified below: 
\begin{itemize}
    \item Player 1: $A = [0,0.8), B = [0.8,0.9), C = [0.9,1]$ 
    \item Player 2: $A = [0,0.1), B = [0.1,0.6), C = [0.6,1]$ 
\end{itemize}
Furthermore, suppose they are shown the object $o_1 = [0.5]$. Then player 1 would announce $A$ and player 2 would announce $B$. Player 1's payoff would be the distance from $o_1$ to the closest boundary of $B$ (player 2's signal) in their memory, (i.e. $|0.5-0.8| = 0.3$), and Player 2's payoff would be the distance from $o_1$ to the closet boundary of $A$ (player 1's signal) in their memory, (i.e. $|0.5-0.1|=0.4$.) Thus, they do not always share payoffs. This is not a partnership game.

Our simulations do not make use of analytical solutions; instead, we fix identical deliberative strategies for both players, which we call the \textit{memory update function}. The memory update function can alternatively be thought of players jointly using numerical approximation towards an equilibrium in a coordination game; deliberative strategies are one possible way for convention to exist without explicit communication \citep{skryms_rational_deliberation}. As we are establishing building blocks for communication, we cannot rely on efficient meta-information sharing so players can coordinate their regions. They must instead establish meaning as conventions. Finally, we motivate the use of a deliberative strategy from our assumption that agents (1) want to retain some priors, and (2) want to align via guessing the other player's future signals. Deliberative strategies are cognitively simpler for agents; they need not process complex analytical solutions on-the-fly. For simplicity, we represent the balance of these two desires with an adaptiveness parameter, $a$, which represents the weight of the new information in comparison to the player's priors at the start of the simulation. 

\noindent
\subsubsection{Semantic Memory}
\begin{itemize}
    \item Each player $p_i$ has a memory $m_i = [c_1^i,c_2^i,c_3^i, \ldots] $ which reference prototypical points that define the Voronoi tessellations representing a player's conceptual memory. Note that each $c_j$ is a vector of $n$-dimensions: $c_j \in [0,1]^n$; the action space is the set of possible values that every $c_j$ can take; however, while players can access full information of their own actions, they cannot access the full representation of the other player's actions. Instead, the information they receive is a signal based on an exogenously drawn object and the corresponding region the other player has assigned it (i.e. a signal \textit{based on} the other player's memory.)
    \item Each player has a set $S^i = [1^i,2^i,3^i, \ldots]$ of finite possible signals, $s \in S$. For simplicity, we assign the index of the prototypical points as the possible signals (i.e. the point $c_3$ will have a signal $3$); consequently, $S_i = S_j$, $\forall i,j$.
    \item Each player derives a memory function from $m_i$ to map to the action space: $M_i(object) = s_{p_i,object}$; this function is a classification algorithm which finds the closest prototypical point $c_s$ to the $object$ and produces $s$ as the signal.
    \item Let the adaptiveness level be set as $a$: if a player wants to update their prior (i.e. a prototypical point) $p_i*$ toward some other point $o_i*$, then they will move $p_i*$ $a\%$ toward $o_i*$. 
    \item This adaptiveness can be degraded overtime by a degradation parameter $d_i*$, so that the new adaptiveness $a_i*'$ is $a_i*' = a_i* \times (1-d_i*)$. Players who have degradeable memory are referred to as non-adaptive players.
\end{itemize}

\subsubsection{The Alignment Game}
\begin{enumerate}
    \item For each player, $p_1$ and $p_2$, generate random points in $n$-dimensions in a $[0,1]^n$ space representing $m_1 = [c_1^1,c_2^1,c_3^1, \ldots] $ and $m_2  = [c_1^2,c_2^2,c_3^2, \ldots] $ . Each player then derives a corresponding classification algorithm $M_1()$ and $M_2()$.
    \item Generate an object, $object$, which is a $n$-dimensional vector representing salient sensory input of an object. \label{new_int_basic}
    \item Generate signals for each player, $s_1$ and $s_2$; where $M_1(object) = s_1$ and $M_2(object) = s_2$. 
    \item If $s_1 = s_2$, return to step \ref{new_int_basic}. Otherwise, proceed to step \ref{update_basic}.
    \item \label{update_basic} Given $s_1 \neq s_2$, update the prototypical points. \begin{enumerate}
        \item Move player 1's prototypical point for the concept signaled by player 2, $c_{s_2}^1$, $a\%$ closer to the $object$. 
        \begin{align*}
            c_{s_2}^1 = movecloser(c_{s_2}^1,object,a)
        \end{align*}
        \item Move player 2's prototypical point for the concept signaled by player 1, $c_{s_1}^2$, $a\%$ closer to the $object$. 
        \begin{align*}
            c_{s_1}^2 = movecloser(c_{s_1}^2,object,a)
        \end{align*}
    \end{enumerate}
    \item Return to Step \ref{new_int_basic} to start new iteration.
\end{enumerate}

\subsubsection{Defining Convergence}
During the game, interlocutors may choose to exit the game if they perceive convergence based on the information they have (i.e. they perceive convergence if they produce the same signal for some set number of consecutive rounds); this may or may not align with actual or true convergence, which is measured by the average distance of the speakers' corresponding prototypical points. We use two different types of convergence to analyze our results from the perspective of Nature (e.g. omnipotent researchers,) as well as the perspective of the players. On a practical level, it is helpful to distinguish between actual and perceived convergence as players in real life may choose to exit the game before actual convergence occurs; we discuss this in the Discussion section.  \\

\noindent
\textbf{Actual Convergence}\label{def:actual_convergence}:
Given some distance tolerance $\delta$, a pair of players' are said to have \textbf{actual convergence} if the average \textbf{conceptual distance} of prototypical points (which refer to corresponding signals) in respective memories, $CD_1$ and $CD_2$ are both less than $\delta$.\\

\noindent
\textbf{Perceived Convergence}\label{def:perceived_convergence}: Given some $\alpha$, players are said to have \textbf{perceived convergence} if their produced signals match for $\alpha$ consecutive rounds. 

\section{Results}
\begin{figure}[H]
    \centering
    \includegraphics[width=\linewidth]{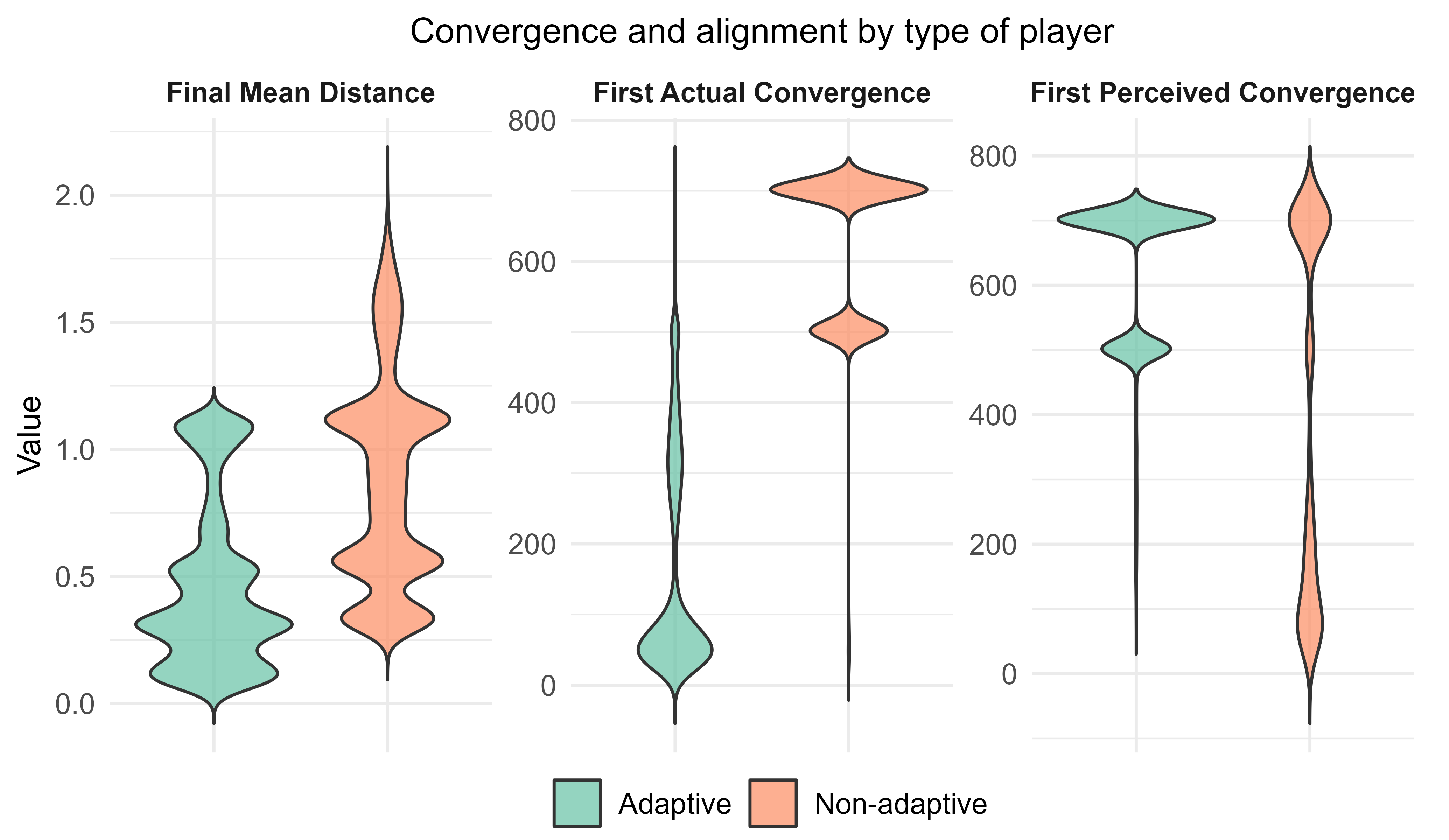}
    \caption{Adaptives are better aligned than non-adaptives but do not perceive this advantage.}
    \label{fig:convergence_by_type}
\end{figure}

\begin{figure}[H]
    \centering
    \includegraphics[width=1\linewidth]{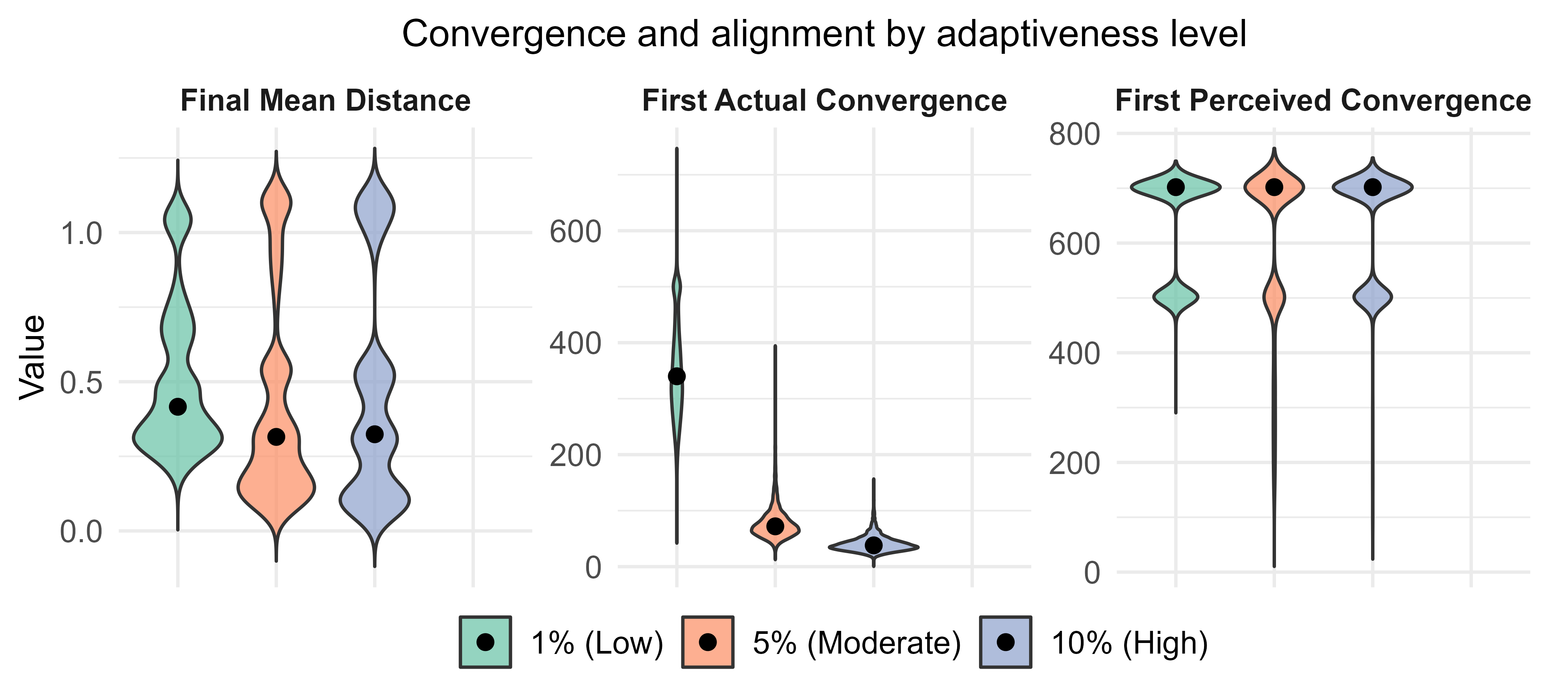}
    \caption{Convergence quality is dependent on adaptability.}
    \label{fig:convergence_by_adaptability}
\end{figure}

\begin{figure}[H]
    \centering
    \includegraphics[width=\linewidth]{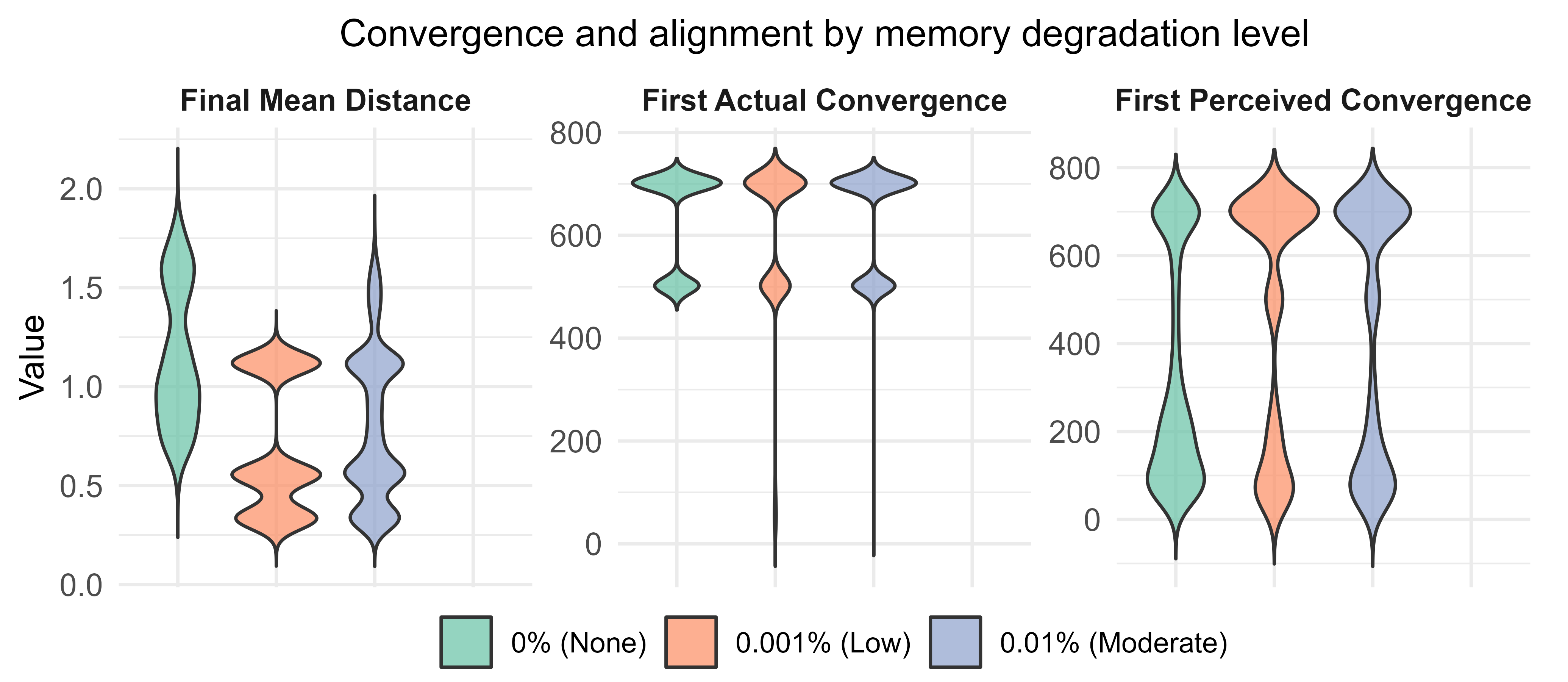}
    \caption{Convergence may or not occur for non-adaptive players; however, how closely aligned they are loosely correlate with final mean distance.}
    \label{fig:converge_nonadaptives}
\end{figure}

We run two sets of counterfactual simulations both groups (i.e. adaptive vs. non-adaptive agents) are shown the same set of randomly generated objects in the same randomly generated order during each simulated play.\footnote{In these presented results, agents only interact with other agents of the same memory type; mixing memory types is an interesting line of research for future work involving network simulations.} We run simulations where agents have 3, 5, or 10 prototypical regions; $1\%$, $5\%$, or $10\%$ adaptiveness levels; and, non-adaptive agents can degrade their memories by $0.001\%$, $0.01\%$, $0.1\%$.\footnote{We set thresholds derived from these possible settings: Our two player results in Figures \ref{fig:convergence_by_type}, \ref{fig:convergence_by_adaptability}, \ref{fig:converge_nonadaptives} are aggregated by a number of parameters that were bootstrapped with a uniform distribution. Moreover, all agents who do not converge are included in this data as max iterations.}
For these simulations, we primarily show results and analysis for (1) the final distances between players' memories, (2) the first instance of \textit{actual convergence}, (3) and the first instance of \textit{perceived convergence}.\footnote{It is important to interpret the mean distances in the context that the overall decision space is a $[0,1]^7$ space; the distance from the origin to the upper bound of the space is about $2.6$: the mean starting distance of our two-player game was about $1.2$.}\footnote{Perceived Convergence threshold was $\# prototypes \times 10$; Maximum iterations (within a conversation) was $500$ (Based on preliminary results, we found that many adaptive players were able to achieve actual convergence around the 200-400 iterations.); the maximum conversations per setting (3, 5 prototypes) was $502$; Maximum conversations per setting ($10$ prototypes) was $702$; Convergence Threshold was $0.075 \times \# prototypes$.} Measures of actual convergence show that overall, adaptive players consistently converge earlier than non-adaptive players; they also end up with overall closer conceptual regions than non-adaptive players. On the other hand, non-adaptive players are more likely to \textit{perceive} convergence earlier.\footnote{Some of the variation in the non-adaptives is due to varying degradation; all are included in Figure~\ref{fig:convergence_by_adaptability} ($0.1\%$, $0.01\%$, $0.001\%$).}\footnote{We also note that all agents who do not converge are included in this data as max iterations which are based on the number of available prototypes; because the thresholds differ across the different agents, the first two figures are designed to highlight the comparative differences across groups and not be literally interpreted.}

We also find that convergence is affected by the adaptiveness level $a$; the final mean distance among 1\% adaptive learners was 0.747, 5\% adaptive learners was 0.63, the final mean distance among 10\% adaptive learners was 0.62.\footnote{More study on this parameter is needed, as it is possible that this decreasing trend may hit an optimal point before more adaptability begins increasing the final mean distance of conceptual spaces in a game.} We do not find substantial differences in the timing of the first perceived convergence; however, we do find that while highly adaptive learners tend to converge first, their prototypical points tend to be comparatively more unstable. We cannot perceive adaptivity gaps between pairings based on external information.

When we look at the rate at which the degradation affects our non-adaptives, the differences are virtually indistinguishable in both measures of actual convergence and perceived convergence; this is likely because the degradation level essentially sets a maximum distance the players are able to move their prototypical points conditional on the objects which are shown to them in the first few hundred simulations. If they are unable to converge before their revisions are negligible, they may not converge, as they are unable to effectively adapt after some number of iterations, regardless of where the other player's prototypical point is. This is shown in the difference between groups in results on final median distance: as the degradation level decreases, the mean final distance decreases.

\begin{figure}[H]
    \centering
    \includegraphics[width=1\linewidth]{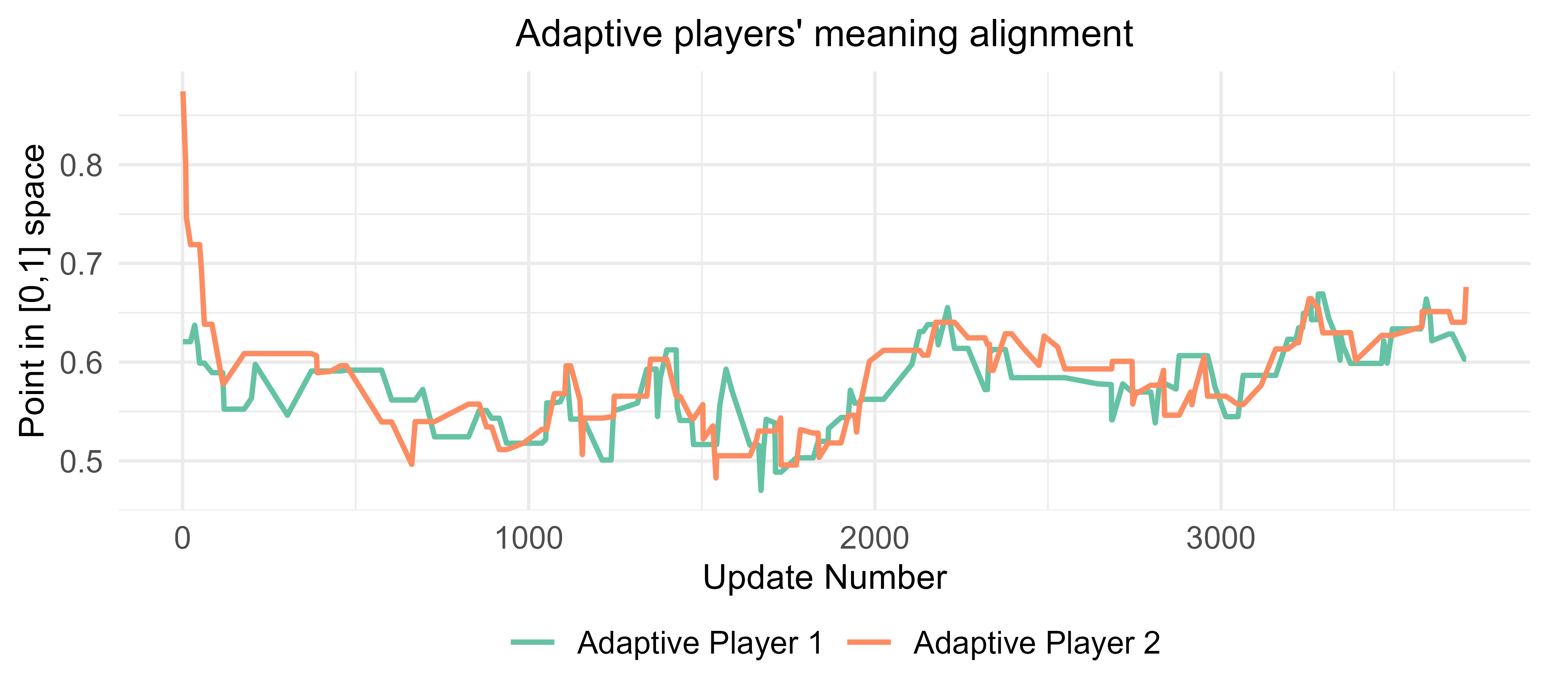}
    \caption{Trajectories for a prototypical point along a dimension for two adaptive players over time}
    \label{fig:adaptive_trajectory}
\end{figure}

\begin{figure}[H]
    \centering
    \includegraphics[width=1\linewidth]{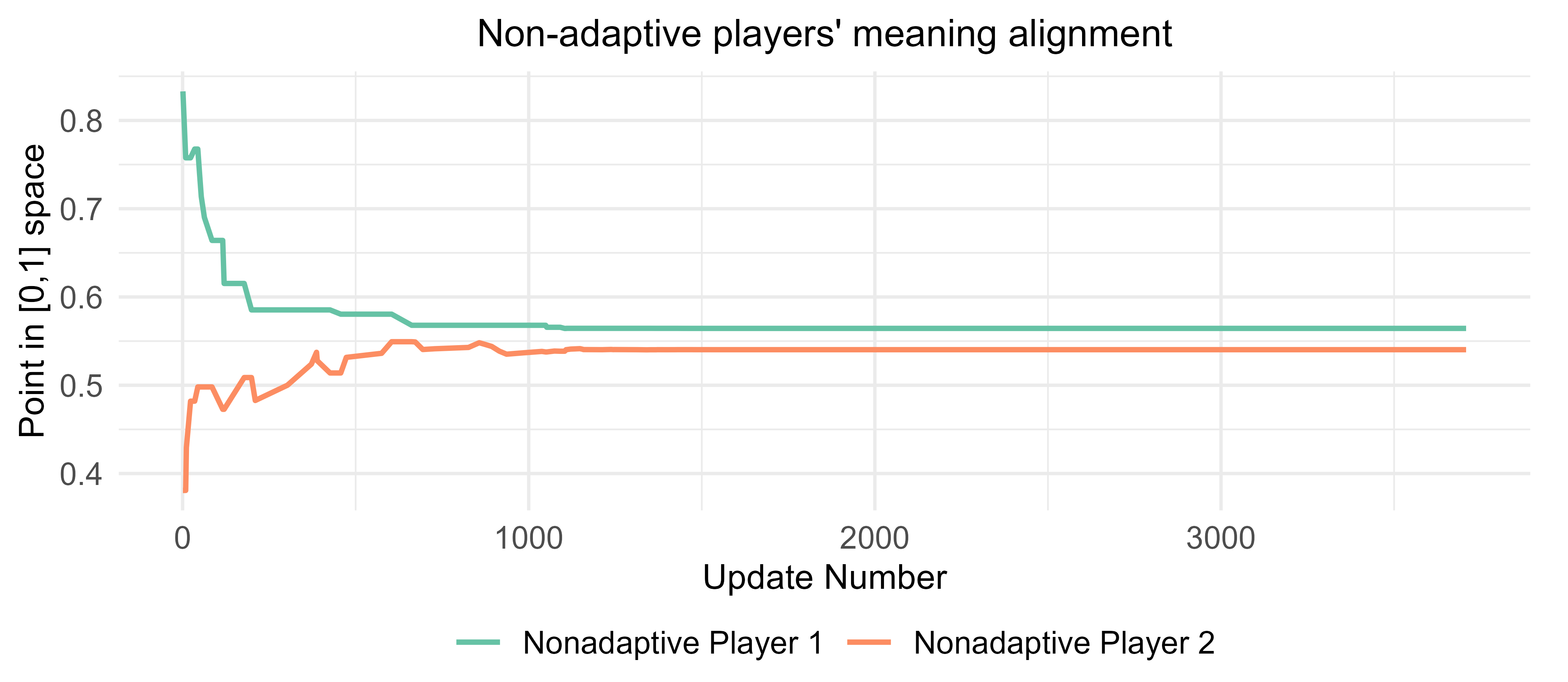}
    \caption{Trajectories for a prototypical point along a dimension for two non-adaptive players over time}
    \label{fig:nonadaptive_trajectory}
\end{figure}

Below, we illustrate the trajectory of a prototype between two adaptive players (Figure~\ref{fig:adaptive_trajectory}) and two nonadaptive players (Figure~ \ref{fig:nonadaptive_trajectory}) over time. Our non-adaptive player decreases their adaptiveness by 1\% each round in comparison to the previous round; consequently, each new data point is able to move the prototypical point increasingly less than the adaptive player as the plays go on.\footnote{ In our simulation, objects are randomly sampled and only one is selected each round, so not all prototypical points each round are updated; additionally, we do not update prototypical points if the agents' signals match; here we only include updates where the agents were shown objects such that the prototypical point referred here was updated.} This dynamic is intuitive given our aggregated results illustrated in figures \ref{fig:convergence_by_type}, and \ref{fig:converge_nonadaptives}. Adaptive players are able to move their points more quickly to one another and converge rather quickly; however, they may not stabilize as they continue updating. In comparison, non-adaptive players are asymptotically stable over time; it is more likely they are able to pass the perceived convergence threshold compared to adaptive players, who may misalign their signals now and then even if they are actually closer to one another on average throughout the simulations or by the end of the simulations.

Measures of actual convergence show that overall, the adaptive players consistently converge earlier than non-adaptive players; they also end up with overall closer conceptual regions than non-adaptive players. On the other hand, non-adaptive players are more likely to \textit{perceive} convergence earlier. Investigation of the trajectory of agents' prototypical points suggest this may in part be due to adaptive players' mischaracterization of instability as misalignment; on the other hand, as non-adaptive players have individually convergent learning strategies, they may mischaracterize how early they have aligned and may need more rounds of successful alignment to actually evaluate when they are aligned. This raises interesting questions for both the viability of these strategies relative to the information environments agents find themselves in. For example, we can imagine that adaptive players may have self-reinforcing strategies; as they continuously feel misaligned to their conversational partners, they may feel like they need to continuously revise their priors in an attempt to realign themselves. This may, however, affect the confidence of these agents in evaluating the reliability of information conveyed using these categories from others. On the other hand, non-adaptive players may also have, given a certain threshold of actual convergence, self-reinforcing strategies, and overestimate the reliability of their interpretations of information from sources they feel aligned to.  

\section{Discussion}
\label{discussion}
 We have shown that the rate which convergence actually occurs and the quality of convergence of the final states are highly dependent on the adaptivity of players in a two player setting. As we take randomized memories, we have shown that differences in convergence can occur without wildly different initial starting points for the players, and without the consideration of network structures. What this tells us is that the level of recency bias we individually have, as well as the recency bias our peers have, can have strong effects on both how we perceive ourselves as members of linguistic communities, as well as our actual understanding of one another. This is motivating evidence for cognitive reasons for not only meaning differences, but the perception of these differences and their emergence. 

Our results may also be interpreted in a more machine-based or semantic memory theoretic context. Theoretically, our results raise additional considerations for when we think about how \enquote{updating} should or can work for vectorized models that learn based on data that is ultimately produced from concepts of actual people. Folding in new data via aggregation without adding recency weights can have the effect, over a course of updates, of a model or agent that is increasingly non-adaptive. While meaning stability may be desireable in certain contexts, we might also consider how the introduction of increasingly non-adaptive agents affect the kind of data that is accessible online that we are learning from. This is especially relevant as we increasingly integrate LLM use in our social media profiles, our personal and work emails, and even in actual news articles and public documents. On the more technical side, it may also be worthwhile to consider the importance of the initial training sets that new models are given- as we model our concept alignment as a two player coordination game, the starting points, particularly for non-adaptive players not only constrain the possible trajectories of their own concepts, but of their conversational partners. We might ask if there are ways to mitigate the effects  the initial training set if we don't have parameters or desiderata for what makes a training set ideal in a given setting, and what data quality concerns we might have going forward given the training sets used in commercialized LLMs.

The work done here thus far has answered the questions of how memory strategies interact with (1) convergence rates (i.e. perceived convergence and actual convergence) and (2) convergence quality. However, these memory \enquote{strategies} were selected a priori as interpretations of possible deliberative strategies for concept emergence. Still, there remain many interesting avenues for future work with the most basic intuitions offered by the Alignment Game.

First, it would be interesting to ask what an \enquote{optimal memory} would be, as there seems to be a trade-off between convergence and adaptability. In our game, we presume that players cooperate given a memory structure; this is reasonable as we are unable to choose the way that semantic memory is stored (and we are also unaware of its structure and function). However, given that our memories have developed in such a way to allow language to function in a way beyond other species, it would be interesting to see if an optimal memory shares similar features to pre-existing models of semantic memory. It would, however, require some definition of how to weigh the desire to not revise one's beliefs, as there is an associated cognitive load, and how to weigh the desire to (1) be aligned to people around us and (2) the desire to be in a network or a clique that is more generally aligned. 

Second, it may also be interesting to allow agents to endogenously choose which other agents to interact with. How should expected costs of convergence play a role in how agents choose whom to speak with? This study could extend along two dimensions: first, we could attempt to simulate a naturally ocurring network structure from social network platforms and compare our results with semantic first-order word association measures (e.g. Reddit); second, we could study the effects of allowing agents to choose who they speak to on convergence. If agents are particularly sensitive to the costs of playing a game with a player they suspect would take many iterations to converge with, they may avoid players who signal that they have drastically different conceptual regions; it would be very difficult then for the network to converge as it may end up fracturing into cliques.

Semantic memory is a topic of both philosophical and practical interest; while we have discussed the implications of research that directly relate to LLMs, understanding how we learn meaning with one another is also important for understanding social issues like misinformation, political extremism and polarization. We find that how memory adapts and how connected individuals are to one another is crucial for how much time interlocutors need to converge in meaning to one another. More research needs to be done theoretically in terms of how memory could have evolutionarily developed as an adaptive function to facilitate a unique feature that humans have: language. 

\section{Conclusion}
 The main aim of this paper is to study how the adaptivity of semantic memory affects the emergence and evolution of conceptual alignment. We presented a novel game characterizing conceptual alignment. Our model is a non-partnership game, allowing a more realistic representation of agent's epistemic access to one another's concepts. For this paper, we focus on using bootstrapped counterfactual simulations of our game to study how rates of memory adaptivity can affect the quality, the speed, and our perceptions of sharing meanings with others. Overall, we find that adaptive players tend to underestimate when they share meaning with other players. We hypothesize that this is because they mistake meaning evolution as misalignment. On the other hand, non-adaptive players tend to overestimate when they are aligned in meaning with others. We hypothesize that this is because they mistake stability of their individual learning strategies for meaning alignment (given some threshold). Our results motivate more work both in terms of how we can think about learning in a semantic memory context, and how societal or real-life incentives and cognitive costs interact. 

\newpage
\printbibliography

\end{document}